\documentclass[preprint]{aastex}

\usepackage{epsfig}

\newcommand{\relemail}{rlingenfelter@ucsd.edu}
\newcommand{\rremail}{ramaty@gsfc.nasa.gov}
\newcommand{\scemail}{scully@gsfc.nasa.gov}
\newcommand{\bzkemail}{benz@wise1.tau.ac.il}

\slugcomment{Submitted to the ApJ}

\def\lsim{\lower.5ex\hbox{$\; \buildrel < \over \sim \;$}}
\def\gsim{\lower.5ex\hbox{$\; \buildrel > \over \sim \;$}}

\begin{document}

\title{Light Element Evolution and Cosmic Ray Energetics}

\author{Reuven Ramaty}
\affil{Laboratory for High Energy Astrophysics\\ NASA/GSFC,
Greenbelt, MD 20771\\ \rremail}

\author{Sean T. Scully\altaffilmark{1}}
\affil{Laboratory for High Energy Astrophysics\\ NASA/GSFC,
Greenbelt, MD 20771\\ \scemail}

\altaffiltext{1}{National Academy of Sciences-National Research
Council Resident Research Associate}

\author{Richard E.Lingenfelter}
\affil{Center for Astrophysics and Space Sciences\\ University of
California at San Diego, LaJolla, CA 92093\\ \relemail}

\and

\author{Benzion Kozlovsky}
\affil{School of Physics and Astronomy, Tel Aviv University,
Israel\\ \bzkemail}

%\begin{abstract}
\noindent ~~~~~~~~~~~~~~~~~~~~~~~~~~~~~~~~~~~~~~~~~~~~~~~~ABSTRACT

\noindent Using cosmic-ray energetics as a discriminator, we investigate the 
viability of evolutionary models for the light elements, Li, Be and 
B (LiBeB). We employ a Monte Carlo code which incorporates hitherto 
ignored effects, the delayed mixing into the ISM both of the 
synthesized Fe, due to its incorporation into high velocity dust 
grains, and of the cosmic-ray produced LiBeB, due to the transport 
of the cosmic rays. We use supernova O and Fe ejecta based on 
calculations and observations, and we normalize the LiBeB production 
to the integral energy imparted to cosmic rays per supernova. We 
find that models in which the cosmic rays are accelerated mainly out 
of the average ISM which is increasingly metal poor at early times, 
significantly under predict the measured Be abundance of the early 
Galaxy, the increase in [O/Fe] with decreasing [Fe/H] indicated by 
recent data notwithstanding. We suggest that this increase could be 
due to the delayed mixing of the Fe. On the other hand, if the 
cosmic-ray metals are accelerated primarily out of supernova ejecta 
enriched superbubbles, such that the cosmic-ray source composition 
as a function of [Fe/H] remains similar to that of the current 
epoch, the measured Be abundances are consistent with a cosmic-ray 
acceleration efficiency that is in very good agreement with the 
current epoch data. This model requires the incorporation of 
neutrino-produced $^{11}$B. We show that, even though the production 
histories of the cosmic-ray produced B and Be and the 
neutrino-produced $^{11}$B are different, B/Be can remain 
essentially constant as a function of [Fe/H]. We also find that 
neither the above cosmic-ray origin models nor a model employing low 
energy cosmic rays originating from the supernovae of only very 
massive progenitors can account for the $^6$Li data at values of 
[Fe/H] below $-$2.

%\end{abstract}


\section{Introduction}

That cosmic-ray driven nucleosynthesis is important to the origin of
the light elements Li, Be and B (LiBeB) has been known for almost
three decades (Reeves, Fowler, \& Hoyle 1970). But only recently was
it realized that the light element themselves, in particular Be
detected in old halo stars formed in the early Galaxy, could provide
new information on cosmic-ray origin, specifically on the source of
the particles that are accelerated to become cosmic rays (Ramaty,
Kozlovsky, \& Lingenfelter 1998).

Although supernova shocks are generally accepted to be the dominant
accelerator of the cosmic rays (at least up to $\sim$10$^5$ GeV),
the source of the particles that are accelerated is still highly
debatable. The first suggestions that cosmic rays are accelerated in
supernova remnants (e.g. Ginzburg \& Syrovatskii 1961; Shapiro 1962)
implicitly assumed that the cosmic-ray source is dominated by fresh
nucleosynthetic material. With the subsequent developments in shock
acceleration theory, it was realized that the most efficient site
for cosmic-ray acceleration is the hot, low density interstellar
medium (see Axford 1981), where the energy loss of the accelerated
particles is minimized and the energy in the supernova shock is not
dissipated by radiative losses. However, a variety of cosmic-ray
injection sources were proposed to explain the large differences
between cosmic-ray and solar abundances. The cosmic-ray source
enrichments were suggested to arise from (i) the atomic
mass-to-charge dependence of supernova shock acceleration (Eichler
1979) in the cooler,  partially ionized phase of the medium, (ii)
the acceleration in the ISM (interstellar medium) of cosmic rays
preaccelerated in stellar coronae, based on the correlation of the
enrichments with first ionization potential (FIP, Cass\'e \& Goret
1978; Meyer 1985), and (iii) the acceleration of grain erosion
products in the average ISM, based on the anti-correlation of the
enrichments with volatility (Epstein 1980). Recent analyses (Meyer,
Drury, \& Ellison 1997) favored a volatility, rather than FIP,
biased cosmic-ray acceleration out of the average ISM, augmented by
a mass-to-charge dependent acceleration of volatiles.

These ideas, prevalent in the 1980's, led to a LiBeB evolutionary
model (hereafter the CRI model) in which the cosmic-ray source
composition at all epochs of Galactic evolution was assumed to be
similar to that of the average ISM at that epoch (Vangioni-Flam et
al. 1990). The excess of the observed Be abundances in low
metallicity stars over the predictions of this model was first
discussed by Pagel (1991), the focus of the discussion being on
whether or not the excess was due to contributions from Big Bang
nucleosynthesis (BBN). The BBN contribution to Be production is now
known (Orito et al. 1997) to be insignificant in comparison with the
Be data at even the lowest metallicities. As a consequence, the CRI
model was modified (Cass\'e, Lehoucq, \& Vangioni-Flam 1995;
Vangioni-Flam et al. 1996; Ramaty, Kozlovsky, \& Lingenfelter 1996;
Vangioni-Flam, Cass\'e, \& Ramaty 1997) by superimposing onto the
cosmic rays accelerated out of the average ISM, a metal enriched
component confined predominantly to low energies (\lsim 100
MeV/nucleon). A strong motivation for this hybrid model was the
report of the detection of C and O nuclear gamma-ray lines from the
Orion star formation region (Bloemen et al. 1994; 1997). These gamma
rays were attributed to low energy cosmic rays (LECRs) highly
enriched in C and O relative to protons and $\alpha$ particles (e.g.
Ramaty 1996), and it was suggested that such enriched LECRs might be
accelerated out of metal-rich winds of massive stars and supernova
ejecta (Bykov \& Bloemen 1994; Ramaty et al. 1996; Parizot, Cass\'e,
\& Vangioni-Flam 1997) by an ensemble of shocks in superbubbles
(Bykov \& Fleishman 1992; Parizot et al. 1997). The Orion gamma-ray
data, however, have now been retracted (Bloemen et al. 1999).
Nonetheless, as the possible existence of the postulated LECRs
remains, new gamma-ray line data are needed to determine the role of
the LECRs in light element production. But one variant of LECR
origin, namely that they are associated with supernovae from only
the most massive stellar progenitors (Vangioni-Flam et al. 1996;
1999), can be ruled out on the grounds of cosmic-ray energetics
(Ramaty \& Lingenfelter 1999 and \S 3.2)

Recent O abundance data, which suggest that [O/Fe] increases with
decreasing [Fe/H] at low metallicities (Israelian et al. 1998,
Boesgaard et al. 1999), led Fields and Olive (1999a) to reexamine
the viability of LiBeB origin based on cosmic-ray acceleration out
of the average ISM (i.e. the CRI model). Hereafter, as is commonly
done in the literature, [O/Fe]$\equiv$log(O/H)-log(O/H)$_\odot$
and [Fe/H]$\equiv$log(Fe/H)-log(Fe/H)$_\odot$, where the ratios of
chemical symbols denote abundance ratios by number. These data
alleviate the shortcomings of the CRI model because the enhanced O
abundance leads to a corresponding enhancement in Be production,
thereby increasing the predicted Be abundance in the early Galaxy.
But this effect by itself is insufficient to render the CRI model
viable (Ramaty \& Lingenfelter 1999). In addition, to account for
the [O/Fe] vs. [Fe/H] data, Fields \& Olive (1999a) allowed the Fe
yields of core-collapse supernovae to take on values constrained
only by these data, rather than by model calculations (Woosley \&
Weaver 1995; Tsujimoto et al. 1995) and observations based on
supernova light curves (see Shigeyama \& Tsujimoto 1998). The
increasing O-to-Fe abundance ratio thus led to decreasing Fe
yields per supernova, which compensated the decreasing Be yield
per supernova in the CRI model. But as was shown (Ramaty \&
Lingenfelter 1999), and will be demonstrated in detail in the
present paper, when supernova Fe yields based on calculations and
observations are used, and if the energy imparted to cosmic rays
per supernova is normalized to reasonable values, the CRI model
still under predicts the Be data by a large factor.

Alternatively, it was suggested (Ramaty et al. 1998; Lingenfelter,
Ramaty, \& Kozlovsky 1998; Higdon, Lingenfelter, \& Ramaty 1998),
that the Be evolution can be best understood in a model (hereafter
CRS) in which the cosmic-ray metals are accelerated out of supernova
ejecta. In this model, the bulk of the Be in the early Galaxy is
produced by accelerated C and O interacting with ambient H and He.
That these ``inverse reactions" are dominant in the early Galaxy was
first suggested by Duncan, Lambert, \& Lemke (1992). Lingenfelter et
al. (1998) and Lingenfelter \& Ramaty (1999) have shown that the
standard arguments (e.g. Meyer et al. 1997; Meyer \& Ellison 1999)
against the supernova ejecta origin of the current epoch cosmic rays
can be answered, and Higdon et al. (1998) and Higdon, Lingenfelter,
\& Ramaty (1999) showed that the most likely scenario is collective
acceleration by successive supernova shocks of ejecta-enriched
matter in the interiors of superbubbles. Due to the fact that
supernova progenitors form mostly in OB associations and have short
lifetimes, the great majority of the core-collapse supernovae go off
in hot, low-density supperbubbles which extend out to hundreds of
parsecs and can last tens of Myr. In addition to the core-collapse
supernovae, a large number of thermonuclear supernovae (Type Ia)
should also occur in these superbubbles due to their large
($\sim$50\%, e.g. Yorke 1986) filling factor, just by chance. Thus,
since 80 - 90 \% of all supernovae are core collapse supernovae with
thermonuclear supernovae making up the remainder (van den Berg \&
Tammann 1991; van den Berg \& McClure 1994), the bulk of the cosmic
rays should be produced by shock acceleration of the metal-enriched
material within supperbubbles. These hot, low density superbubbles
are, in fact, the ``hot phase" of the interstellar medium where
shock acceleration of cosmic rays is expected (e.g. Axford 1981;
Bykov \& Fleishman 1992) to be most effective because the energy
losses of the accelerated particles are minimized and the supernova
shocks do not suffer major radiative losses, as they would in a
denser medium. This scenario of collective acceleration by
successive supernova shocks in the hot ISM is consistent with the
delay between nucleosynthesis and acceleration (time scales of $\sim
10^5$ yr), suggested by the recent $^{59}$Co and $^{59}$Ni
observations (Binns et al. 1999; Wiedenbeck et al. 1999).

In the present paper we describe in detail a new Monte-Carlo
approach to LiBeB evolution in which the concept of the
nucleosynthetic yields and energy in cosmic rays of individual
supernovae is explicitly built in. A Monte-Carlo approach to
Galactic chemical evolution was previously employed by Copi
(1997), but not for $^6$Li, Be and B. Copi's (1997) aim was the
investigation of the expected scatter of the calculated
abundances. Scatter in the abundance ratios of low metallicity
stars, in a model in which abundance patterns are determined by
individual supernovae, was considered by Tsujimoto, Shigeyama, \&
Yoshii (1999), and the evolution of the light elements in this
model was treated by Suzuki, Yoshii, \& Kajino (1999). We employ
the Monte-Carlo simulation as a tool for solving the equations of
Galactic chemical evolution, one that allows us to investigate
hitherto ignored effects, such as the delayed mixing into the ISM
both of the synthesized Fe, due to its incorporation into high
velocity dust grains, and of the cosmic-ray produced light
elements, due to the transport of the cosmic rays. Preliminary
presentations of our approach were given in Ramaty, Lingenfelter,
\& Kozlovsky (1999a,b) and Ramaty \& Lingenfelter (1999). We also
incorporate light element yields calculated per supernova for a
variety of metallicity-dependent cosmic-ray and ISM abundances,
employing a previously developed code (Ramaty et al. 1997). We
first demonstrate the energetic inconsistency of the CRI model,
including the recent variants by Fields \& Olive (1999a) and Suzuki
et al. (1999), namely light element production by cosmic-rays
accelerated out of the average ISM. Next we treat the evolution of
B and examine the conditions than can lead to a constant
evolutionary B-to-Be abundance ratio when $^{11}$B production by
neutrinos in supernovae is included. Finally, we consider the
evolution of $^6$Li and the accompanying $^7$Li, and examine the
viability both of the CRI model (Fields \& Olive 1999b) and the
model employing LECRs from only the most massive supernova stellar
progenitors (Vangioni-Flam et al. 1999) to account for the $^6$Li
abundance data reported at very low metallicities (Hobbs \&
Thorburn 1997; Smith, Lambert, \& Nissen 1998; Nissen et al.
1999).

\section{Fe and O Production and Evolution}

We consider a one-zone model using a Monte-Carlo code. We accumulate
primordial gas according to the prescription
\begin{eqnarray}
{d{\rm M_{ISM}}(t) \over dt} = {\rm M_{\rm h} \over \tau_{\rm h}
\big[1 - e^{-{\rm T_g}/\tau_{\rm h}}\big]}
e^{-t/\tau_{\rm h}} + {\rm M_{\rm d} \over \tau_{\rm d}^2
\big[1-(1+{\rm T}_{\rm g}/\tau_{\rm d})
e^{-{\rm T}_{\rm g}/\tau_{\rm d}}\big]} t e^{-t/\tau_{\rm d}}~~,
\end{eqnarray}
\noindent where the first and second terms on the right hand side
represent mass accumulation into the halo and disk, respectively.
This approach is similar to that of Chiappini et al. (1999), where
the halo and disk are formed from external gas characterized by
different infall time scales, $\tau_{\rm h}$ and $\tau_{\rm d}$,
respectively. For the formation of the disk, we have adopted the
infall rate suggested by Kobayashi et al. (1998) which has an
extra $t$ multiplying the exponential. The time scale for the halo
formation is thought to be less the 1 Gyr while that for the disk
is significantly longer (e.g. Chiappini et al. 1999; Boissier \&
Prantzos 1999). We remove ISM mass by star formation, taking the
star formation rate proportional to the ISM mass, $\Psi=\alpha{\rm
M_{ISM}}$ where $\alpha$ is generally found to be on the order of
several tenths of Gyr$^{-1}$ (e.g. Prantzos, Cass\'e, \&
Vangioni-Flam 1993; Mihara \& Takahara 1996; Kobayashi et al.
1998). We set up a grid of 55 logarithmically spaced time bins,
ranging from 1 Myr to 15 Gyr, and accumulate mass into these bins
by the above prescription. Then for each bin we generate an
ensemble of stellar masses normalized to the assumed star
formation rate and distributed according to the Salpeter IMF in
the range 0.1 to 100 M$_\odot$. Starting from the first bin and
advancing in time, we calculate the removed mass and the returned
mass for each star in the ensemble, the latter depending on
stellar lifetimes and the remnant mass.

Galactic Fe and O production is due to the $>$10 M$_\odot$ stars
which explode as core collapse supernovae. An additional source of
Fe is provided by thermonuclear supernovae (Type Ia) whose
contributions become significant at times later than about 1 Gyr.
For the core collapse supernovae, in each time bin and for each star
in the ensemble, we calculate the ejected O and Fe masses for two
cases. In the first case (hereafter WW95), which we chose in order
to maximize both the Be and O productions relative to that of Fe, we
take the minimum Ni-Fe yield models of Woosley \& Weaver (1995),
which practically vanish at progenitor masses of 30 M$_\odot$ and
above. Specifically, we use the $Z = 10^{-4} Z_\odot$ models
corresponding to the lowest ejecta kinetic energy ($W_{\rm SN}$) at
30 M$_\odot$, and those at intermediate $W_{\rm SN}$ at 35 and 40
M$_\odot$, because these still yield vanishing Ni-Fe yields but
finite O yields. As no calculations for progenitors more massive
than 40 M$_\odot$ are available in Woosley \& Weaver (1995), we
assume that the ejected O and Fe masses for such progenitors are
equal to those of 40 M$_\odot$.

The second case (hereafter TS) is unbiased, based on the yields
given by Tsujimoto \& Shigeyama (1998). These were obtained
(Shigeyama \& Tsujimoto 1998) by combining the observed [Mg/Fe]
vs. [Mg/H] data with nucleosynthetic calculations (Woosley \&
Weaver 1995; Tsujimoto et al. 1995) of Mg yields for various
progenitor masses, and an inferred relationship between these
masses and [Mg/H] obtained by assuming that at low metallicities
the observed abundance ratios are determined by the contribution
of individual supernovae. These TS yields are in fairly good
agreement with $^{56}$Ni masses derived from current epoch
supernova light curves (see figure 4 in Shigeyama \& Tsujimoto
1998). Furthermore, for the progenitor mass range 30 to 40
M$_\odot$, they are bracketed by the Woosley \& Weaver (1995)
yields for ejecta kinetic energy, $W_{\rm SN}$, ranging from about
1 to 3$\times$10$^{51}$ erg.

We summarize the ejected masses for both the WW95 and TS cases in
Table~1. Even though the WW95 yields depend on the W$_{\rm SN}$,
they are within about 50\% of 1.5$\times$10$^{51}$ erg, so we
assume this value for the entire progenitor mass range from 10 to
100 M$_\odot$ in both the WW95 and TS cases. We employ these
ejected masses at all metallicities, justified by the fact that
the IMF averaged O-to-Fe ratio in the Woosley \& Weaver (1995)
calculations varies by less than 25\% over the entire metallicity
range from $10^{-4} Z_\odot$ to  $Z_\odot$.

The contributions of the Type Ia supernovae, resulting from
accreting white dwarfs, was modeled by scaling their Fe yield to
that of the core collapse supernovae, using a factor of
proportionality that varies with [Fe/H] (Kobayashi et al. 1998).
This variation results from the fact that the ``planetary nebula"
winds of low mass stars that blow off the outer layers of the
stars to form the white dwarfs, only become effective at
metallicities [Fe/H] $> -1$, so that there is negligible white
dwarf formation before that, as Kobayashi et al. (1998) have
modeled. Using a star formation rate coefficient $\alpha = 0.5$
and $\tau_{\rm d} = 5 {\rm Gyr}$, we adjusted the variation to
obtain a good fit to the [O/Fe] vs. [Fe/H] data, yielding a factor
that for the WW95 case ranges from 2.5 at [Fe/H] = 0 to 1.2 at
[Fe/H] = $-1$, and vanishes for [Fe/H] $< -1$. For the TS case,
the factor is 1.7 at [Fe/H]=0 and 0.8 at [Fe/H] = $-1$, and again
vanishes at lower [Fe/H]. These values are consistent with a ratio
of about 10 between Fe yields of thermonuclear and core collapse
supernovae (Nomoto et al. 1997; Woosley \& Weaver 1995), and with
the current epoch ratio of about 0.1 to 0.25 between the
frequencies of these supernovae (van den Berg \& Tammann 1991; van
den Berg \& McClure 1994). The calculated O-to-Fe abundance ratio
([O/Fe]$\equiv$log(O/Fe)$-$log(O/Fe)$_\odot$), together with
recent data (Israelian et al. 1998; Boesgaard et al. 1999), are
shown in Figure~1a, where we see that for both the WW95 and TS
cases the data are reasonable well fit for [Fe/H] greater than
about $-1.5$. At lower values of [Fe/H], WW95 still provides a
possible fit, but TS is inconsistent with the data because of the
large Fe yields of the massive stars in this case (Table~1).

We suggest that the observed increase in [O/Fe] at low [Fe/H]
could be caused by delays in the deposition of the synthesized
products into the star forming regions of the ISM due to
differences in transport and mixing. In the simulation we choose
the mixing times randomly in the interval 0$-\tau({\rm mix})$ and
allow for the possibility that $\tau_{\rm Fe}$(mix)$>$$\tau_{\rm
O}$(mix). We choose a short mixing time for oxygen, $\tau_{\rm
O}$(mix)=1 Myr, because we would expect the bulk of the O and
other volatiles in the ejecta to mix with the ISM after the remnant
slows down to local sound speeds of the $\sim$300 km s$^{-1}$ or
less, which occurs in $\sim 3\times10^5$ yr or less, even in the
tenuous hot phase of the ISM where most core collapse supernovae
occur (see Higdon et al. 1998 and the references therein). But we
consider longer mixing times for Fe, assuming that the bulk of the
ejected Fe would be incorporated into high velocity refractory
dust grains which continue moving for longer periods of time
before they stop and can be incorporated into newly forming stars.
The incorporation of a large fraction of the synthesized Fe into
dust grains is supported both by observations (Kozasa, Hasegawa \&
Nomoto 1991) suggesting the massive condensation of iron oxide and
other refractory grains at velocities $\simeq$2,500 km s$^{-1}$ in
the supernova 1987A and by observations (Naya et al. 1996) of the
Galactic 1.809 MeV gamma-ray line resulting from the decay of
$^{26}$Al, most likely produced in Type II supernovae (e.g.
Woosley \& Weaver 1995). The observed width of the $^{26}$Al line
implies that the radioactive aluminum is still moving at
velocities $>$450 km s$^{-1}$ some 10$^6$ yrs after its formation
and long after the associated supernova remnants have slowed to
thermal velocities, suggesting that the bulk of the synthesized
$^{26}$Al is in high velocity dust grains which take a much longer
time to slow down. Since both aluminum and iron form highly
refractory compounds, it is likely that the bulk of the
synthesized Fe also resides in such grains which could travel much
farther before they came to rest than the volatiles trapped in the
plasma which are more rapidly slowed by the swept up gas and
magnetic fields. Only about 20\% of the O is bound up in such
refractory grains (see Lingenfelter et al. 1998). Since most of
the supernovae occur in the hot, tenuous superbubble phase of the
ISM with densities as low as 10$^{-3}$ H cm$^{-3}$, a typical
supernova grain of diameter 10$^{-5}$ cm, or a column depth of
10$^{19}$ H cm$^{-2}$, will lose roughly $1/n$ of its
momentum in traversing and sweeping up $n$ column depths of ISM.
Thus, such grains could travel for path lengths as great as 30
kpc, and times as long as 100 Myr in slowing down from 2500 to 300
km s$^{-1}$. The rectilinear distances, of course, can be
considerable less because the grains will carry some net charge
and will propagate like high rigidity particles in the
interstellar magnetic field.

The effects of such a delay between the mean deposition times into
the ISM of refractories, such as Fe, and volatiles such as O, are
shown in Figure~1b for a possible $\tau_{\rm Fe}$(mix) of 30 Myr.
We see that with this delay, both the WW95 and TS cases become
consistent with the data, showing that delayed Fe deposition could
indeed be the cause for the rise of [O/Fe] with decreasing [Fe/H]
at very low metallicities. In this connection, it is interesting
to note that, unlike [O/Fe], the abundance ratios of the
$\alpha$-nuclei Mg, Si, Ca and Ti relative to Fe do not increase
with decreasing [Fe/H] below [Fe/H] $= -1$ (Ryan, Norris, \& Beers
1996). This may be consistent with the fact that these elements
are also refractory, and thus are affected by mixing in the same
way as is Fe. Our suggestion that the bulk of the synthesized Fe
is incorporated into high velocity dust grains, and thus widely
distributed throughout the halo, differs from the scenario
proposed by Tsujimoto et al. (1999), who assumed that the
nucleosynthetic products of each supernova, including Fe, only
enrich the shell swept up by that supernova. Indeed, Tsujimoto et
al. (1999) do not provide an explanation for the rise of [O/Fe]
vs. [Fe/H].

The value of $\tau_{\rm h}$ influences the efficiency of the
mixing in producing a rising [O/Fe] with decreasing [Fe/H]. This
can be seen in Figure~2a, where [O/Fe] does not increase as much
for $\tau_{\rm h} = 1 {\rm Gyr}$ as it does for $\tau_{\rm h} = 10
{\rm Myr}$, because for the longer halo formation time the 30 Myr
delay is clearly less significant. In the subsequent calculation,
we shall only use $\tau_{\rm h} = 10 {\rm Myr}$. As a consistency
check, we show in Figure~2b the measured [O/H] vs. [Fe/H]
(Israelian et al. 1998; Boesgaard et al. 1999) together with our
calculated evolutionary curve which provides an acceptable fit, as
can be seen. The linear fit in Figure~2b will be used in the
subsequent calculations of light element production (\S 3).
Finally, in Figure~3 we show the temporal evolution of [Fe/H] for
the two cases (WW95 and TS) and for two Fe mixing times. We see
that higher values of [Fe/H] are achieved with the TS case, which
corresponds to higher ejected Fe masses (Table~1). In addition, as
expected, the delayed Fe deposition causes [Fe/H] to be lower at
early times. Our age-metallicity results are quite consistent with
other calculations (e.g. Mihara \& Takahara 1996) and data (see
Pagel 1997, fig. 8.13).

We also note that [C/Fe], unlike [O/Fe], does not increase with
decreasing [Fe/H] below [Fe/H]$= -1$. As shown by Timmes, Woosley and
Weaver (1995), this results from the fact that at early times both
the C and Fe come primarily from core collapse supernovae of massive
stars. At later times, [Fe/H]$>-$1, the ratio is still roughly
constant because of the compensating effects of the increased C
contribution from the winds of intermediate mass stars and the
increased Fe contribution from the thermonuclear supernovae of the
white dwarf remnants of such stars. Thus, we take [C/Fe] as a
constant at the solar value, which is quite consistent with the
measured values in stars with [Fe/H] ranging from $-$3 to 0.5 (e.g.
Timmes et al. 1995).

\section{Light Element Production and Evolution}

The light element production by accelerated particles per supernova
depends on several factors (e.g. Ramaty et al. 1997): the
composition of both the ambient medium and the accelerated
particles, the energy spectrum of the accelerated particles, the
energy per supernova imparted to the accelerated particles, and the
interaction model for the accelerated particles, characterized by a
path length for escape from the Galaxy, X$_{\rm esc}$.

We summarize the assumed abundances in Table~2. We take the ambient
medium (ISM) metal abundances at a given [Fe/H] to be solar, scaled
with 10$^{\rm [Fe/H]}$, except for O for which the scaling is given
by 10$^{\rm 0.63[Fe/H]}$ (Figure~2b), and for the $\alpha$ elements,
for which the scaling is increased by a factor of 3 at [Fe/H] $<
-1$, consistent with the evolution of these elements (e.g. Ryan et
al. 1996), to adjust for the more recent effects of $\alpha$ element
poor thermonuclear supernovae. For consistency with Big Bang
nucleosynthesis, we scale the He abundance with a slowly varying
function, ranging from 1 at [Fe/H]=0 to 0.75 at the lowest
metallicity. We consider two compositions for the accelerated
particles. For the CRS model (\S 1) we take a constant cosmic-ray
source model which approximates the acceleration of particles
primarily out of supernova ejecta enriched superbubbles (Higdon et
al. 1998; 1999). In this model the abundances are independent of
[Fe/H] and set equal to those given by Lund (1989) and Engelmann et
al. (1990) for the current epoch cosmic rays, consistent with recent
ACE results (Lijowski et al. 1999). For the CRI model we take an
evolving cosmic ray source model which assumes the acceleration of
particles primarily out of the interstellar medium with selective
enrichments (Meyer et al. 1997). The abundances for the CRI model
are the same as the CRS at [Fe/H]=0. But at lower metallicities, the
CRI abundances are taken to be proportional to those of the ambient
medium, with the abundances of the volatile elements, C, N, O, Ne,
and S, enhanced by factors of 1.5, 1.8, 2, 2.5, and 5, respectively,
consistent with a mass-to-charge dependent acceleration of volatiles
(Ellison et al. 1997), and those of the refractory elements, Mg, Si
and Fe, enhanced by a factor of 20, consistent with the current
epoch cosmic-ray source to solar abundance ratio (Meyer et al. 1997)
and a grain sputtering injection. Ne and heavier elements do not
contribute to LiBeB production, but they are included in the
calculation of the cosmic-ray energy content.

The accelerated particle source energy spectra are given by an
expression appropriate for shock acceleration,
\begin{eqnarray}
Q_{\rm cr}(E) \propto
{p^{-s} \over \beta} {\rm e}^{-E/E_0},
\end{eqnarray}
where $p, c\beta$ and $E$ are particle momentum/nucleon, velocity
and energy/nucleon, respectively, and $E_0$ is a turnover energy
that we set at a large value (e.g. 10 GeV/nucleon) for the CRS and
CRI models, implying a spectrum extending to ultrarelativistic
energies. For both these models we take $s=2.5$. For consistency
with the recent calculations of Vangioni-Flam et al. (1999, \S 3.2),
for the LECR model we also use eq.(2), but with much smaller value
of $E_0$ (e.g. 30 MeV/nucleon) and $s=2$.

We first calculate $Q/W$, the total number of nuclei $Q$ produced by
an accelerated particle distribution normalized to the integral
cosmic-ray energy $W$, for a given source energy spectrum and
composition, and interacting in an ambient medium of given
composition. $Q/W$ is obtained independent of the evolutionary
calculation, as the only evolutionary input is via the abundances
that are given as a functions of [Fe/H], as described above.
Furthermore, we make the reasonable assumption that the accelerated
particle source energy spectrum is independent of [Fe/H]. We employ
the light element production code that was described in detail by
Ramaty et al. (1997). The resultant CRS and CRI $Q/W$'s for Be and
$^6$Li are shown in Figure~4 as functions of [Fe/H], for ${\rm
X}_{\rm esc} = 10~{\rm g cm}^{-2}$ typical of currently inferred
values (e.g. Engelmann et al. 1990) for ``leaky box" cosmic-ray
propagation models, except for Be in the CRI model, for which we
also show $Q/W$ for ${\rm X}_{\rm esc} \rightarrow \infty$,
representing the ``closed Galaxy" model which maximizes $Q/W$. As
can be seen, the assumption of a closed Galaxy increases $Q/W$ by
only a factor of about 2. If we had taken the variation of C similar
to that of O (see Table~2), the CRI Q/W would have increased by only
30\%, but as we have pointed out in \S 2, it is more appropriate to
scale [C/O] linearly with [Fe/H]. While $Q(^6{\rm Li})/W$ is not
very different for the CRS and CRI models, $Q({\rm Be})/W$ is
drastically different for the two models, reflecting the fact that
efficient Be production in the early Galaxy can only result from C
and O enriched accelerated particles, as is expected if the cosmic
rays are accelerated from supernova ejecta enriched matter (i.e. the
CRS composition). In the CRI model, $^6$Li in the early Galaxy is
produced almost entirely in interactions of $\alpha$-particles with
He, so that the difference between the CRS and CRI curves for $^6$Li
is due to the contributions of interactions involving the C and O.
We do not show the closed Galaxy $Q({\rm Be})/W$ beyond
[Fe/H]$=-0.5$, as this model is not appropriate for the current
epoch cosmic rays.

The evolutionary code described in the previous section requires the
light element yields per supernova. These are given by $(Q/W) W_{\rm
cr}$, where $W_{\rm cr} = \eta W_{\rm SN}$, $\eta$ being the
acceleration efficiency. We assume that $W_{\rm
SN}$=1.5$\times$10$^{51}$ erg is the same for both the core collapse
and thermonuclear supernovae, and that it is independent of
progenitor mass, as discussed in \S 2, as well as of [Fe/H].
Likewise, we assume that $\eta$ is a constant, which we determine by
normalizing the calculations to the data. Since for the CRI model
the accelerated particles originate from the average ISM, we allow
all supernovae, core collapse and thermonuclear, to produce Be and
$^6$Li. The situation is somewhat different for the CRS model. As
discussed in \S 1, about 85\% of the core collapse and 50\% of the
thermonuclear supernovae are expected to occur in the superbubble
hot phase (HISM) of the ISM. For these supernovae we assume that the
light elements are produced according to the $Q/W$ for the CRS
model. For the remaining 15\% core collapse and 50\% thermonuclear
supernovae, the production is according to the $Q/W$ of the CRI
model. Implicit in this prescription is the assumption that the
cosmic-ray acceleration efficiency $\eta$ is the same for supernovae
occurring in the HISM and in the average ISM. This overestimates the
CRI contribution, since shock acceleration studies (e.g. Axford
1981; Bykov \& Fleishman 1992) suggest that the acceleration
efficiency should be much higher in the superbubble HISM than in the
average ISM, but even this CRI is not significant in the early
Galaxy. We delay the light element deposition into the ISM because
of the finite propagation and interaction time of the accelerated
particles. The delay depends on the mean ISM density $n_{\rm H}$
encountered by the cosmic rays. Using the light element production
code (Ramaty et al. 1997), we derived the appropriate distributions,
which for a mean ISM density of 1 atom cm$^{-3}$ correspond to an
average delay of several Myr for both the CRS and CRI models.

\subsection{Beryllium and Boron}

Using the metallicity dependent $Q/W$'s, we calculated the evolution
of the Be and B abundances. The results for Be are shown in
Figures~5a and 5b, for the WW95 and TS cases, respectively, together
with data from Garcia Lopez (1999) and Boesgaard \& King (1993), the
latter being limited to disk stars. We fix the acceleration
efficiency $\eta$ by normalizing the CRS calculations to the data,
yielding the energy in cosmic rays per supernova, $W_{\rm cr} = \eta
W_{\rm SN}$, given in the figures. The fact that this cosmic-ray
energy of 1.5$\times$10$^{50}$ erg is in excellent agreement with
the required energy in cosmic rays per supernova (e.g. Lingenfelter
1992), based on current epoch cosmic-ray data and supernova
statistics, and the fact that the corresponding acceleration
efficiency $\eta$ of 10\% is quite reasonable for shock acceleration
(e.g. Axford 1981), both provide strong support for the validity of
the CRS model. On the other hand, the same W$_{\rm cr}$ for the CRI
model leads to Be abundances that under predict the data by about
two orders of magnitude at the lowest [Fe/H], thereby demonstrating
the energetic inconsistency of this model. Our result differs from
that of Fields \& Olive (1999a) who concluded, for their best set of
parameters, that the excess energy per supernova needed at [Fe/H]=
$-3$ only exceeds the current value by about a factor of 5. Most of
the discrepancy (see Ramaty \& Lingenfelter 1999 and \S 1 for more
detail) is probably caused by differences in the employed Fe ejected
masses, which are not quoted by Fields \& Olive (1999a). Since the
values for the masses that we use are much more tightly constrained
by both nucleosynthetic calculations and supernova light curves (\S
2), we believe that the discrepancy between the predictions of the
CRI model and the data that we are demonstrating is indeed real. The
possible use of a closed Galaxy model (Figure~4), which was not
employed by Fields \& Olive (1999a), would increase the predicted CRI
values by only a factor of 2. In addition, the shapes of the CRS
curves are also in much better agreement with the data than those of
the CRI model.

The effect of the Be deposition delay due to cosmic-ray transport
can be clearly seen in Figures~5 below [Fe/H]$\simeq$$-2$. As
pointed out above, for $n_{\rm H}=1~{\rm cm}^{-3}$, the delay is on
the order of a few million years and consequently for $n_{\rm H}=0.1
{\rm cm}^{-3}$ (a possible value for a spherical halo of radius 10
kpc and containing 10$^{10}$ M$_\odot$ of gas), it is on the order
of tens of millions of years. The results of Figure~5, showing
significant differences between the $n_{\rm H}=1~{\rm cm}^{-3}$ and
$0.1~{\rm cm}^{-3}$ curves up to about [Fe/H]=$-$2, can be
understood from the age-metallicity relations of Figure~3, which
indicate that [Fe/H]$\simeq$$-2$ is achieved in a time period (about
100 Myr) of the same order as the delay for $n_{\rm H}=0.1~{\rm
cm}^{-3}$. The delayed deposition of Be was also considered by
Suzuki et al. (1999) who suggested that this delay would cause a
large scatter in log(Be/H) vs. [Fe/H] at low [Fe/H]. But this
scatter should be much smaller if the bulk of the Fe is in high
velocity dust grains, because both the Fe and the Be will be widely
distributed throughout the halo. The Suzuki et al. model is
effectively a mix of CRS (2\%) and CRI (98\%) cosmic rays. But this
model fails on two major points. First, with a source consisting of
only 2\% (CRS) supernova ejecta, the Be production efficiency is
correspondingly reduced at low metallicities ([Fe/H] $\sim -$3) and
would thus require roughly 50 times as much energy per supernova as
the full CRS model (Figure 5), or 5 times the total ejecta energy.
Second, the dominance of the CRI cosmic-ray component at [Fe/H]
$>-1$, not shown by Suzuki et al. (1999), would lead to the very
significant overproduction of Be.

The evolution of the cosmic-ray produced $^{11}$B and $^{10}$B is
essentially identical to that of Be because both production ratios,
$Q({\rm B})/Q({\rm Be}) \simeq 14$ and $Q(^{11}{\rm B})/Q(^{10}{\rm
B}) \simeq 2.4$, are practically independent of [Fe/H]. We note in
figure~2a of Ramaty et al. (1997) that a constant $Q({\rm B})/Q({\rm
Be})$ for the CRS model is achieved only when the two-step
processes, including secondary interactions discussed in that paper,
are taken into account. As the isotopic ratio of 2.4 is
significantly lower than that measured both in meteorites
(4.05$\pm$0.2, Chaussidon \& Robert 1995) and the current epoch ISM
(3.4$\pm$0.7, Lambert et al. 1998), we augment the $^{11}$B
production with the contribution of core collapse supernovae where
the $^{11}$B results from carbon spallation by neutrinos (Woosley et
al. 1990; Woosley \& Weaver 1995). In the evolutionary calculation
we allow only the core collapse supernovae to contribute
$\nu$-produced $^{11}$B, and we employ the metallicity dependent
$^{11}$B yields of Woosley \& Weaver (1995) which, unlike those of O
and Fe, vary by about a factor of 2 from $10^{-4} Z_\odot$ to
$Z_\odot$. We then normalize these $\nu$-produced yields with a
factor $f_\nu$, which we take to be independent of [Fe/H], and
adjust $f_\nu$ to yield the meteoritic boron isotope ratio of 4.05
at [Fe/H]=0. As B is volatile, we delay the deposition of the
$\nu$-produced $^{11}$B with the same short mixing time (1 Myr) that
we used for oxygen, but the delay for the cosmic-ray produced B is
n$_{\rm H}$-dependent, similar to the Be deposition delay. The
results for B/Be and $^{11}$B/$^{10}$B are shown for $f_\nu= 0.28$
in Figures~6a and 6b, respectively, together with data, Duncan et
al. (1997) and Garcia Lopez et al. (1998) for B/Be, and Chaussidon
\& Robert (1995) and Lambert et al. (1998) for $^{11}$B/$^{10}$B.
The rise in both ratios below [Fe/H]$\simeq$$-2$ for n$_{\rm H}=
0.1~$cm$^{-3}$ results from the delayed deposition of the cosmic-ray
produced B and Be (tens of Myr for n$_{\rm H}= 0.1~$cm$^{-3}$). We
see that for both values of n$_{\rm H}$ the calculations are
consistent with the data, although future, more precise B/Be data
could distinguish between the models. Our calculated constant value
of log(B/Be) = 1.35 for n$_{\rm H}= 1$cm$^{-3}$ is in very good
agreement with the measured mean of 1.27$\pm$0.47 (Garcia Lopez et
al. 1998), and only slightly larger than the mean given by Duncan et
al. (1997, see comment in Garcia Lopez et al. 1998 concerning the
difference between the two sets of data). This calculated B/Be,
combined with the cosmic-ray produced B/Be=14, implies that 38\% of
the boron is $\nu$-produced in core collapse supernovae.

Our essentially constant B/Be and $^{11}$B/$^{10}$B contrast with
the results of Vangioni-Flam et al. (1996) whose B/Be and
$^{11}$B/$^{10}$B evolutionary curves, with $\nu$-produced $^{11}$B
included, exhibited prominent maxima around [Fe/H]$\simeq$$-2$. This
led Duncan et al. (1997) to argue against significant $\nu$
production, as the B/Be data did not show evidence for such a
maximum. The variation found by Vangioni-Flam et al. (1996) for
[Fe/H] below the maximum was the consequence of the assumption that
the accelerated particles that produce the Be and B in the early
Galaxy originate only from core collapse supernovae having very
massive ($>60 M_{\odot}$)stellar progenitors, while the
$\nu$-produced $^{11}$B results from all core collapse supernovae.
It is the difference in lifetimes of the progenitors that lead to
the rise below [Fe/H]$\simeq$$-2$. But this model involving only
very massive progenitors encounters serious energetic difficulties
(Ramaty \& Lingenfelter 1999 and \S 3.2). The decline for [Fe/H]
above the maximum in the Vangioni-Flam et al. (1996) model was due
to the increasingly significant CRI contribution which increases the
cosmic-ray produced Be and B and thus lowers the ratios. Such a
decrease would have been present in our calculations as well,
because of the increasing $Q/W$ (Figure~4) and the contributions of
the thermonuclear supernovae above [Fe/H]$\simeq$$-1$. But these
increases are compensated for by the increasing $\nu$-produced
$^{11}$Be yields, as discussed above. Thus, while $\nu$-produced
$^{11}$B is needed to account for the high B isotopic ratio, it is
not in conflict with an essentially constant B/Be as a function of
[Fe/H].

\subsection{Lithium}

The calculated $^6$Li evolutionary curves are shown in Figure~7 for
the CRS and CRI models. We included primordial $^6$Li adopting an
abundance ($^6$Li/H)$_{\rm BBN}$=5$\times$10$^{-14}$, which is the
mean value of the commonly quoted range (e.g. Olive \& Fields 1999).
While the uncertainties in the main $^6$Li producing cross section
are quite large, the extreme upper limit of around 10$^{-12}$
(Nollet, Lemoine, \& Schramm 1997) seems rather unlikely (see
Vangioni-Flam et al. 1999). The calculated CRS and CRI curves in
Figure~7 are thus almost totally cosmic-ray produced. Unlike those
for Be, they are not very different, reflecting the dominant
contribution of the $\alpha$-particle induced reactions on He which
gives a $Q/W$ that is almost independent of [Fe/H] (see Figure~4).
We also show a total Li evolutionary curve, consisting of a Big Bang
contribution and cosmic-ray produced Li. As the cosmic-ray $^7$Li to
$^6$Li production ratio, $Q(^7{\rm Li})/Q(^6{\rm Li}) \simeq 1.6$,
is also practically independent of [Fe/H] (Ramaty et al. 1997), we
evaluated the $^7$Li evolution by simply scaling ($^7$Li)$_{\rm
CRS}$ to $^6$Li, and by adding primordial $^7$Li with abundance
$^7$Li/H=1.8$\times$10$^{-10}$ (Molaro 1999). The resultant
evolutionary curve is consistent with the data (compiled by M.
Lemoine, private communication 1997), except at values of [Fe/H]
near zero and above where a variety of additional Galactic source
contribute (e.g. Romano 1999).

Considering the calculations and data for $^6$Li, we first note
the good agreement with the meteoritic data, providing further
support for the validity of the procedure that we have adopted,
namely the normalization of light element production to a
cosmic-ray energy per supernova that is consistent with the
corresponding current epoch value. Concerning the rest of the
data, we see that both the CRS and CRI curves fail to account for
the measured $^6$Li data at [Fe/H]$<$$-$2. This result, specifically for
the CRI model, is again in conflict with Fields \& Olive (1999b)
who claimed that their evolutionary calculation can
simultaneously account for these and the meteoritic data. As for
Be, we believe that the difference is in the adopted Fe yields per
supernova, as we have discussed above. Both the CRS and CRI models
over predict some of the $^6$Li upper limits at higher [Fe/H], in
particular that at [Fe/H]=$-$1 for the star HD 134169. But this
apparent discrepancy could be due to $^6$Li depletion on this star
(see Cayrel 1999).

Another explanation for the high $^6$Li abundances at [Fe/H]=$-$2.4
and $-$2.2 was suggested by Vangioni-Flam et al. (1999). As
mentioned above, they consider a model in which a new component of
energetic particles is accelerated exclusively from the ejecta of
supernovae originating from very massive progenitors ($>$50
M$_\odot$). The accelerated particles in this model are confined to
low energies (i.e. $E_0$=10 or 30 MeV/nucleon in eq.[2]), and their
composition is assumed to vary with [Fe/H]. Vangioni-Flam et al.
(1999) point out that the $^6$Li-to-Be production ratio varies
inversely with the accelerated particle C/O, a fact that combined
with an increasing C/O could provide an explanation for the higher
$^6$Li/Be below [Fe/H]=$-$2 than at [Fe/H]=0. Specifically, for the
early Galaxy they adopt the composition of a supernova from a 40
M$_\odot$ progenitor (the U40B model of Woosley \& Weaver 1995) for
which C/O=0.09 as opposed to the CRS value which is an order of
magnitude higher (Table~2).

To investigate the validity of this model we have evaluated $Q/W$ as 
a function of [Fe/H] for $E_0=30$ MeV/nucleon and $s =2$ (eq.[2]), 
the parameters employed by Vangioni-Flam et al. (1999). For the 
ambient medium we used the composition as given in Table~2. We 
employed 3 compositions for the accelerated particles, CRS 
(Table~2), CRS(metal) taken to be identical to CRS but with 
vanishing H and He abundances, and U40B. The results are shown in 
Figure~8 for both $^6$Li and Be. We see that the $^6$Li-to-Be 
production ratio can indeed decrease, changing from about 15 for the 
U40B composition to about 3 for that of CRS(metal). But the 
variation is almost entirely due to the increased $Q({\rm Be})/W$, 
with $Q(^6{\rm Li})/W$ being essentially independent of composition. 
But when the U40B case is compared with CRS (Figure~8), $Q(^6{\rm 
Li})/Q({\rm Be})$ remains essentially constant. We thus evaluated 
the $^6$Li evolution using the $Q/W$ for the U40B composition up to 
[Fe/H]=$-$1, and transitioning to that for CRS(metal) at higher 
[Fe/H]. We allowed only core collapse supernovae from $>$50 
M$_\odot$ progenitors to contribute, as modeled by Vangioni-Flam et 
al. (1999), we normalized the energy, $W_{\rm cr} = \eta W_{\rm 
SN}$, that each such supernova imparts to accelerated particles to 
fit the $^6$Li data below [Fe/H]=$-$2, and we ignored the 
thermonuclear supernovae. The results are shown in Figure~9, where 
we see that an unreasonably large energy is required, 
5$\times$10$^{51}$ ergs per supernova, mostly because only a small 
number of supernovae are allowed to contribute to the acceleration 
of the low energy cosmic rays. When compared to the ejecta kinetic 
energy of 2$\times$10$^{51}$ erg of the U40B model (Woosley \& 
Weaver 1995), the acceleration efficiency, $\eta$, is 250\%. This 
finding is a very strong argument against the validity of this 
model. For E$_0 = 10$MeV/nucleon, another possibility suggested by 
Vangioni-Flam et al. (1999) the required energy is even larger, by 
about a factor of 3. Another difficulty is that for [Fe/H]$>$$-$1.5 
the model predicts too much $^6$Li and associated $^7$Li, keeping in 
mind that this contribution from the hypothetical low energy 
cosmic-ray component must be added to that of the standard cosmic 
rays (see Vangioni-Flam et al. 1996). This would lead to a conflict 
both with the total Li and the meteoritic $^6$Li data. We believe 
that the difference between our result and that presented by 
Vangioni-Flam et al. (1999), whose $^6$Li evolutionary curve is 
consistent with both the early Galactic and meteoritic data, is due 
to the fact that they normalized their $^6$Li production to that of 
the Be rather than keeping the energy per supernova constant as we 
do. But since there is no justification for assuming that the energy 
imparted to accelerated particles per supernova is metallicity 
dependent, we conclude that this is another argument against the 
validity of this model.

It thus appears that the measured $^6$Li abundances below 
[Fe/H]=$-$2 require that $^6$Li be produced in the early Galaxy 
without significant accompanying Be, a requirement naturally 
satisfied by $\alpha$-particle interaction with He, and that this 
production be limited to early times only. To consider the required 
energy in cosmic rays, we calculate $Q(^6{\rm Li})/W$ for the CRI 
composition at [Fe/H]=$-4,$ which ensures that the accompanying Be 
production will be negligible. $Q(^6{\rm Li})/W$ is shown in 
Figure~10 as a function of $E_0$ (eq.[2]), as the energy spectrum of 
these hypothetical cosmic rays is of course not known. Assuming the 
maximal production efficiency of about 0.1 atoms/erg, we find that 
10$^{-10}$ erg per H atom are needed to produce the required 
$^6$Li/H of $\sim$10$^{-11}$. We defer to future research the 
analysis of the various early Galactic or extra-Galactic scenarios 
that could satisfy this requirement.  For example, if the process 
operates in the Galactic halo of mass 10$^{10}$M$_\odot$, then about 
10$^{57}$ ergs are needed.

\section{Discussion and Conclusions}

Our first conclusion concerns the viability of CRI model for the
origin of the cosmic-ray produced LiBeB. In this model, often
referred to as ``the standard cosmic-ray model", at all epochs of
Galactic evolution the cosmic rays are assumed to be accelerated out
of the average ISM which is increasingly metal poor at early times.
Employing core collapse supernova Fe ejecta based on both
calculations and observations, we have shown that the increase in
[O/Fe] with decreasing [Fe/H] notwithstanding, this model
significantly under predicts the measured Be abundances at low
metallicities, contrary to the conclusion of Fields \& Olive (1999a).
We also point out that the scenario proposed by Suzuki et al.
(1999), in which CRI cosmic-ray production of Be and B in the early
Galaxy is supplemented by that of a small (2\%) component of
supernova ejecta cosmic rays, fails as well because of the
energetics. To emphasize these results, we outline here a simple
argument that demonstrates the issue:

The IMF averaged Fe yield per core collapse supernova is about 0.1
M$_\odot$ for both the WW95 and TS cases (\S 2), based on
calculations and observations. To reach [Fe/H]=$-$3 for these
models, at least 20 Myr are required (Figure~3), a time period equal
to or longer than the lifetimes of all core collapse supernova
progenitors, justifying IMF averaging over progenitor masses $>$10
M$_\odot$. The essentially linear dependence of log(Be/H) vs. [Fe/H]
(Figure~5) implies a constant Be/Fe$\simeq$10$^{-6}$. This, coupled
with the Fe yield per supernova, implies an essentially constant Be
yield per supernova, $\simeq$2$\times$10$^{48}$ atoms. Using the
relevant CRI $Q/W$ $\simeq$2$\times$10$^{-4}$ atoms/erg (Figure~4),
for evaluating the Be abundance at [Fe/H]=$-$3, we obtain the
required energy per supernova of $\simeq$10$^{52}$ erg, which is
what rules out the CRI model. On the other hand, for the CRS model
(cosmic-ray metal acceleration predominantly out of relatively fresh
nucleosynthetic material in the superbubble hot phase of the ISM),
the same Be yield per supernova, combined with the CRS $Q/W$,
$\simeq$0.02 atoms/erg (Figure~4), implies 10$^{50}$ ergs per
supernova, in excellent agreement with the current epoch value.

Both the CRS and CRI models under predict the B isotope ratio
measured in meteorites. Thus, we assume that 38\% of the boron is
neutrino-produced $^{11}$B in core collapse supernovae, which leads
to an evolutionary $^{11}$B/$^{10}$B that is consistent with the
meteoritic data. The fact that the calculated neutrino-produced
$^{11}$B yield increases with metallicity (\S 3.1) compensates for
the increase in the cosmic-ray produced B and Be near [Fe/H]=0,
caused both by the proton and $\alpha$-particle induced reactions
and the contributions of the thermonuclear supernovae, thus leading
to an essentially constant B/Be.

The CRS model, while providing a good fit to the meteoritic $^6$Li
abundance, appears to under predict the measured $^6$Li/H below
[Fe/H]=$-$2 (\S 3.2). The discrepancy between these data and the
predictions of the CRI model are even larger. We showed, that a
model that postulates low energy cosmic rays from supernovae
originating from only very massive progenitors is energetically
untenable. This, however, does not rule out the existence of a low
energy cosmic-ray component of less restrictive origin which may
play a role in light element origin (see Ramaty \& Lingenfelter
1999, table 2, for calculations of the required energy per
supernova). Nevertheless, the origin of the early Galactic $^6$Li
abundances remains unclear. We show that the required energetics are
at least 10$^{-10}$ erg per H atom.

We also provide a possible explanation for the increase in [O/Fe]
with decreasing [Fe/H], in the delayed mixing of the
nucleosynthesized Fe due to its incorporation into high velocity
dust grains. A consequence of this scenario is an essentially
constant abundance ratio of the refractory $\alpha$-nuclei (Mg, Si,
Ca and Ti) relative to Fe at [Fe/H] $< -1$ where core collapse
supernovae dominate, as opposed to an increasing O-to-Fe abundance
with decreasing [Fe/H]. Current data are consistent with such
constancy.

\acknowledgments

We acknowledge support from NASA's Astrophysics Theory Program.

%\eject

\begin{figure}[t]
  \begin{center}
    \leavevmode
\epsfxsize=10.cm \epsfbox{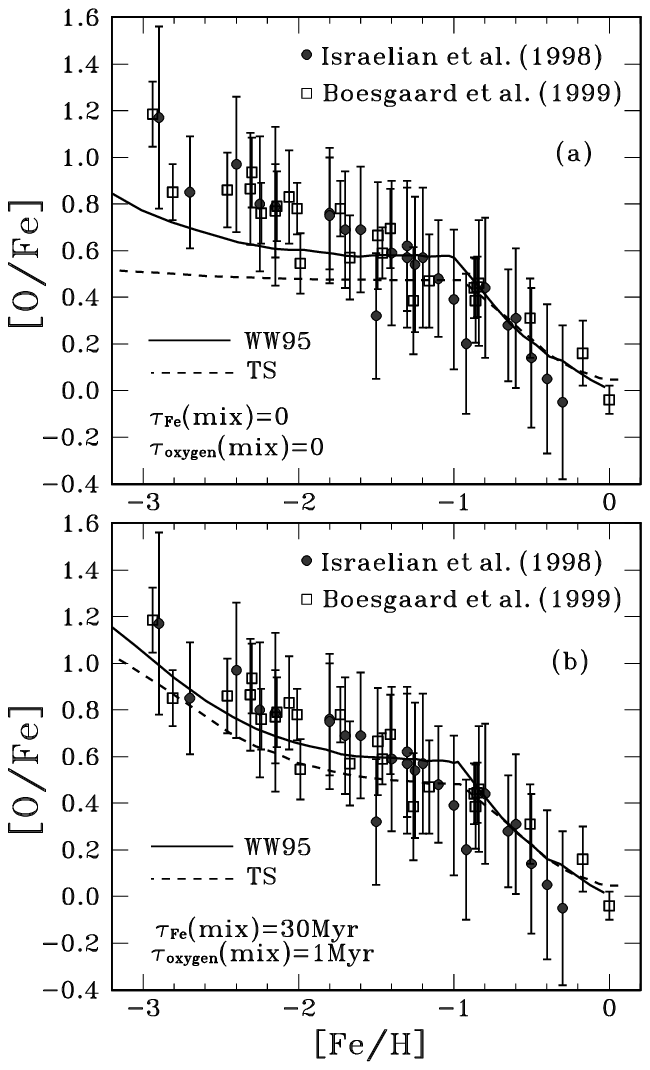}
  \end{center}
  \caption{Galactic evolution of the O-to-Fe abundance ratio
vs. [Fe/H] for supernova O and Fe yields taken form Woosely \& Weaver
(1995, WW95) and Tsujimoto \& Shigeyama (1998, TS),
assuming zero and finite Fe and O mixing delay times,
panels (a) and (b), respectively. In both panels, the characteristic
halo and disk infall times are $\tau_{\rm h}=10 {\rm Myr}$ and
$\tau_{\rm d}=5 {\rm Gyr}$, the star formation rate coefficient
$\alpha= 0.5 {\rm Gyr}^{-1}$, and the ratio of halo-to-disk masses
M$_{\rm h}$/M$_{\rm d}$=0.1.}
\end{figure}

\eject

\begin{figure}[t]
  \begin{center}
    \leavevmode
\epsfxsize=10.cm \epsfbox{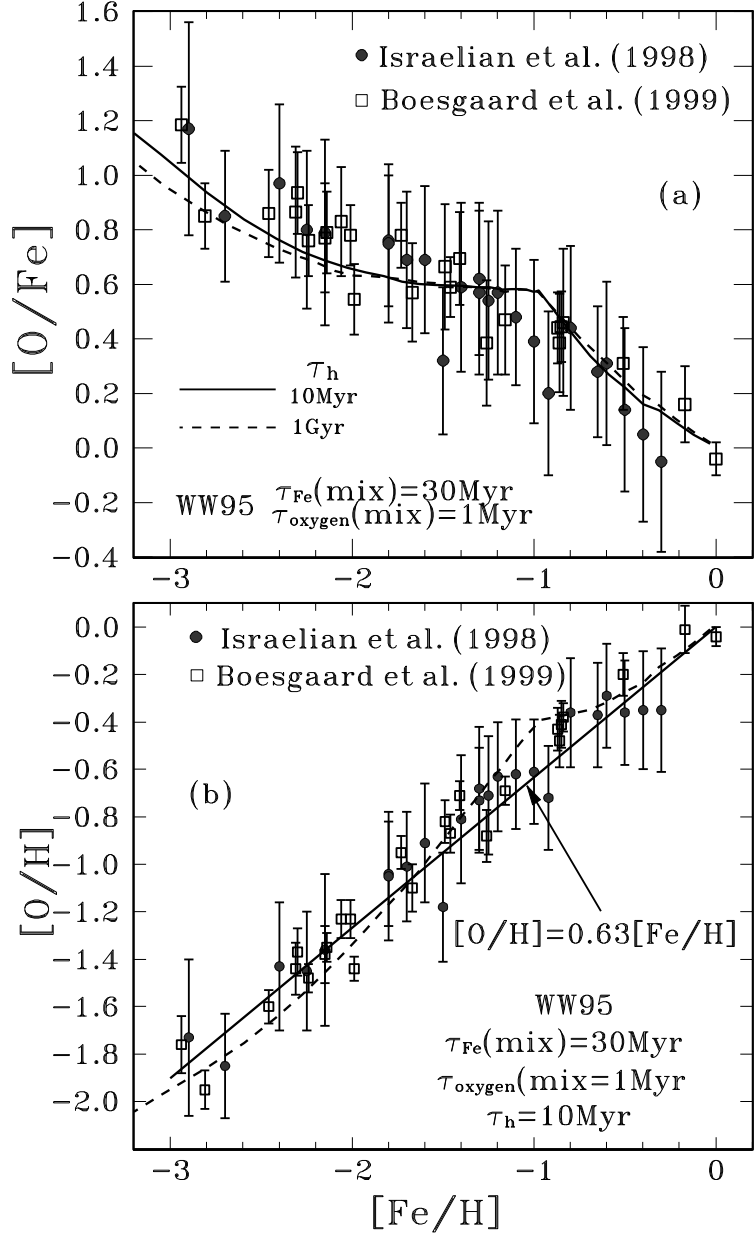}
  \end{center}
  \caption{Galactic evolution of: (Panel a) the O-to-Fe
abundance ratio vs. [Fe/H] for two values of the characteristic
halo infall time $\tau_{\rm h}$; (Panel b) the O-to-H ratio
vs. [Fe/H] for $\tau_{\rm h}$ = 10Myr, compared to the empirical
fit [O/H]=0.63[Fe/H]. These calculations use only the WW95 yields
while the other evolutionary parameters are the same as those in Fig. 1.}
\end{figure}

\begin{figure}[t]
  \begin{center}
    \leavevmode
\epsfxsize=15.cm \epsfbox{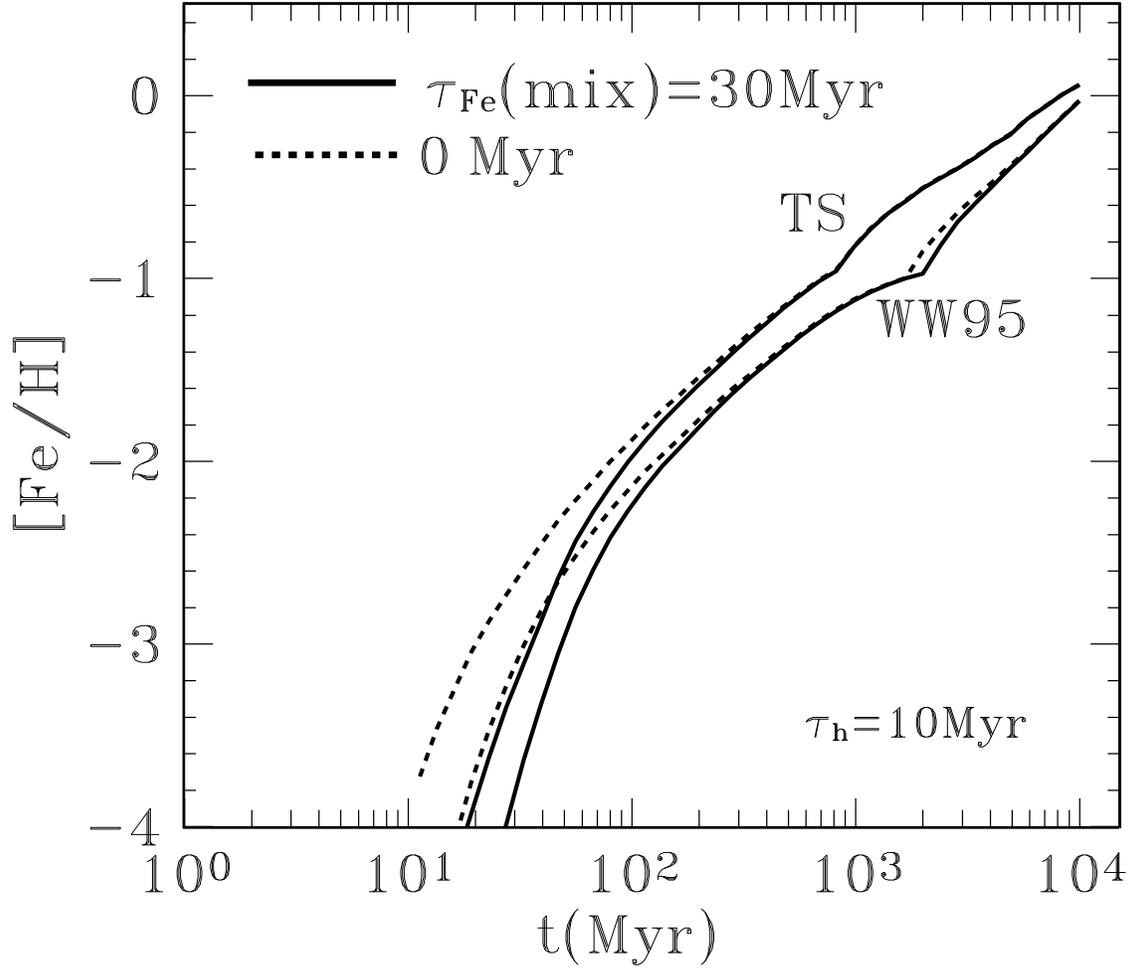}
  \end{center}
  \caption{Fe abundance as a function of time for the two assumed
  ejecta cases (Table~1), and two Fe mixing times.}
\end{figure}

\begin{figure}[t]
  \begin{center}
    \leavevmode
\epsfxsize=15.cm \epsfbox{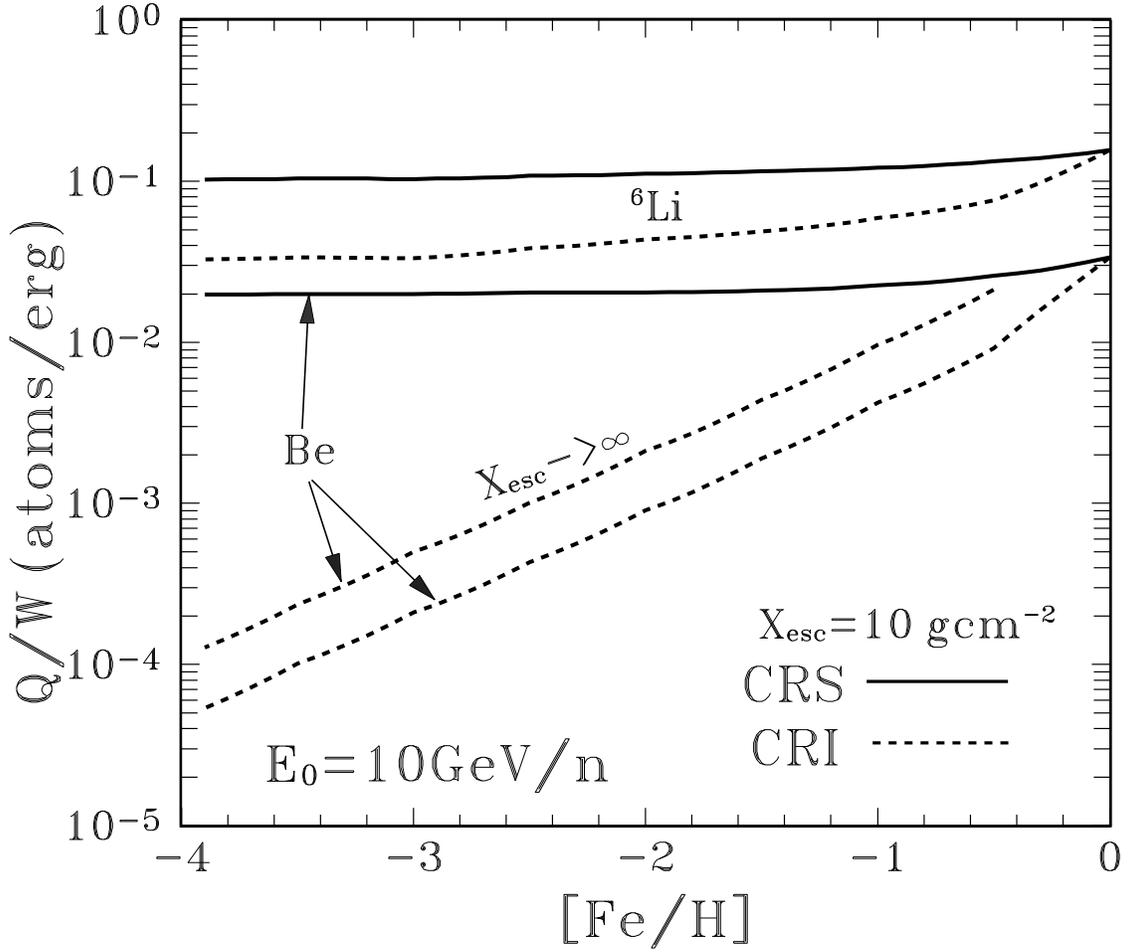}
  \end{center}
  \caption{Number of Be and $^6$Li atoms produced per unit
cosmic-ray energy for models of the cosmic ray abundances derived
from supernova ejecta enriched superbubbles CRS and the
average interstellar medium CRI, using the current epoch
cosmic-ray escape path length from the Galaxy with X$_{\rm esc}$ = 10 g
cm$^{-2}$ and an infinite path length representing a possible
early ``closed Galaxy" which increases Q/W by about a factor of 2.}
\end{figure}

\begin{figure}[t]
  \begin{center}
    \leavevmode
\epsfxsize=10cm \epsfbox{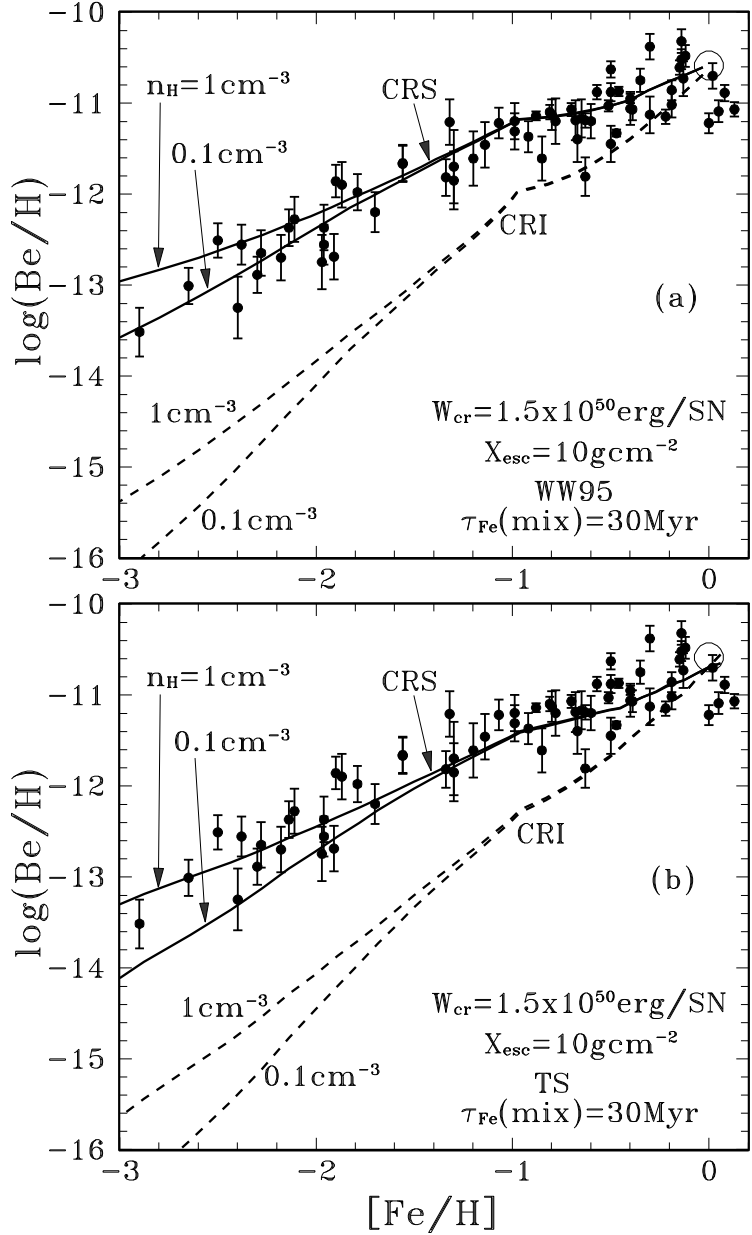}
  \end{center}
  \caption{Be abundance evolution for two cases (WW95 and TS,
Table~1) of core collapse supernova Fe and O ejecta, panels (a)
and (b), respectively. The CRS and CRI curves
correspond to cosmic-ray acceleration out of supernova
ejecta and average ISM, respectively. All calculations are for a
cosmic-ray energy spectrum extending to ultrarelativistic
energies. n$_{\rm H}$ is mean density of the ambient medium
encountered by cosmic rays, which determines the delay of the
deposition of the synthesized Be.}
\end{figure}

\begin{figure}[t]
  \begin{center}
    \leavevmode
\epsfxsize=10.cm \epsfbox{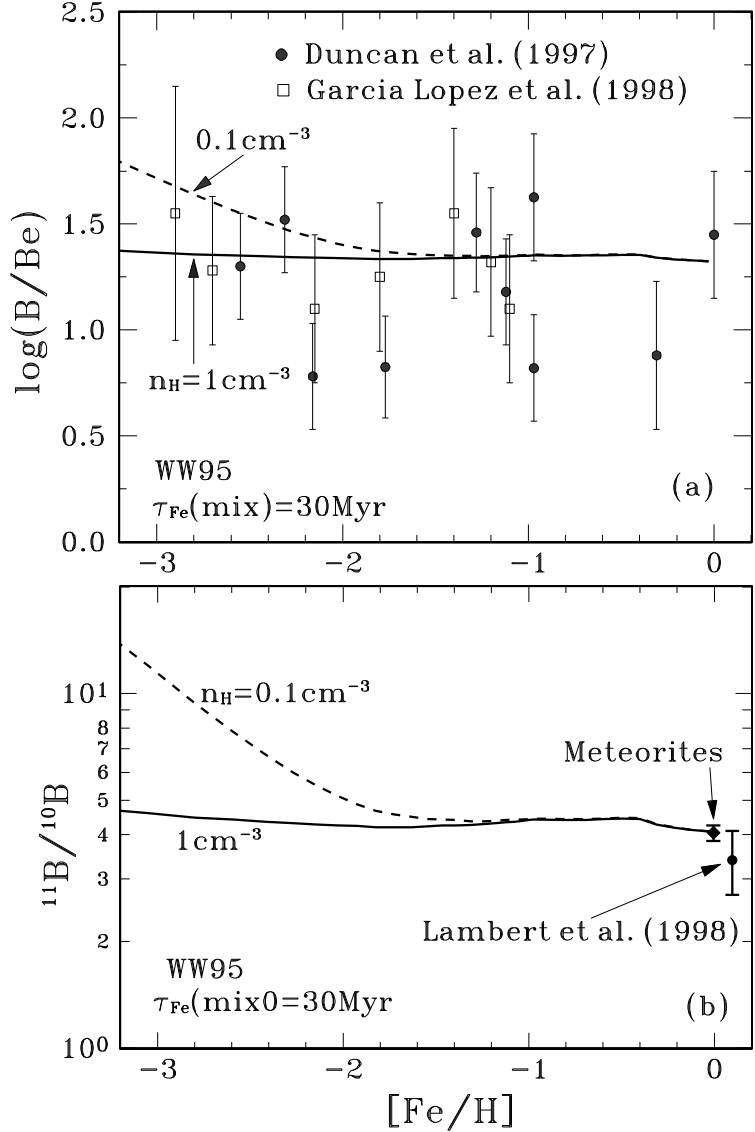}
  \end{center}
  \caption{B-to-Be and $^{11}$B-to-$^{10}$ abundance ratios vs.
  [Fe/H], panels (a) and (b), respectively, with the contribution
  of $\nu$-produced $^{11}$B included. The B isotope data for
  meteorites are from Chaussidon \& Robert (1995). n$_{\rm H}$ is
  mean density of the ambient medium encountered by cosmic rays,
  which determines the delay of the
  depositions of the cosmic-ray produced Be and B. The assumed
  delay for the $\nu$-produced $^{11}$B is 1 Myr.}
\end{figure}

\begin{figure}[t]
  \begin{center}
    \leavevmode
\epsfxsize=15.cm \epsfbox{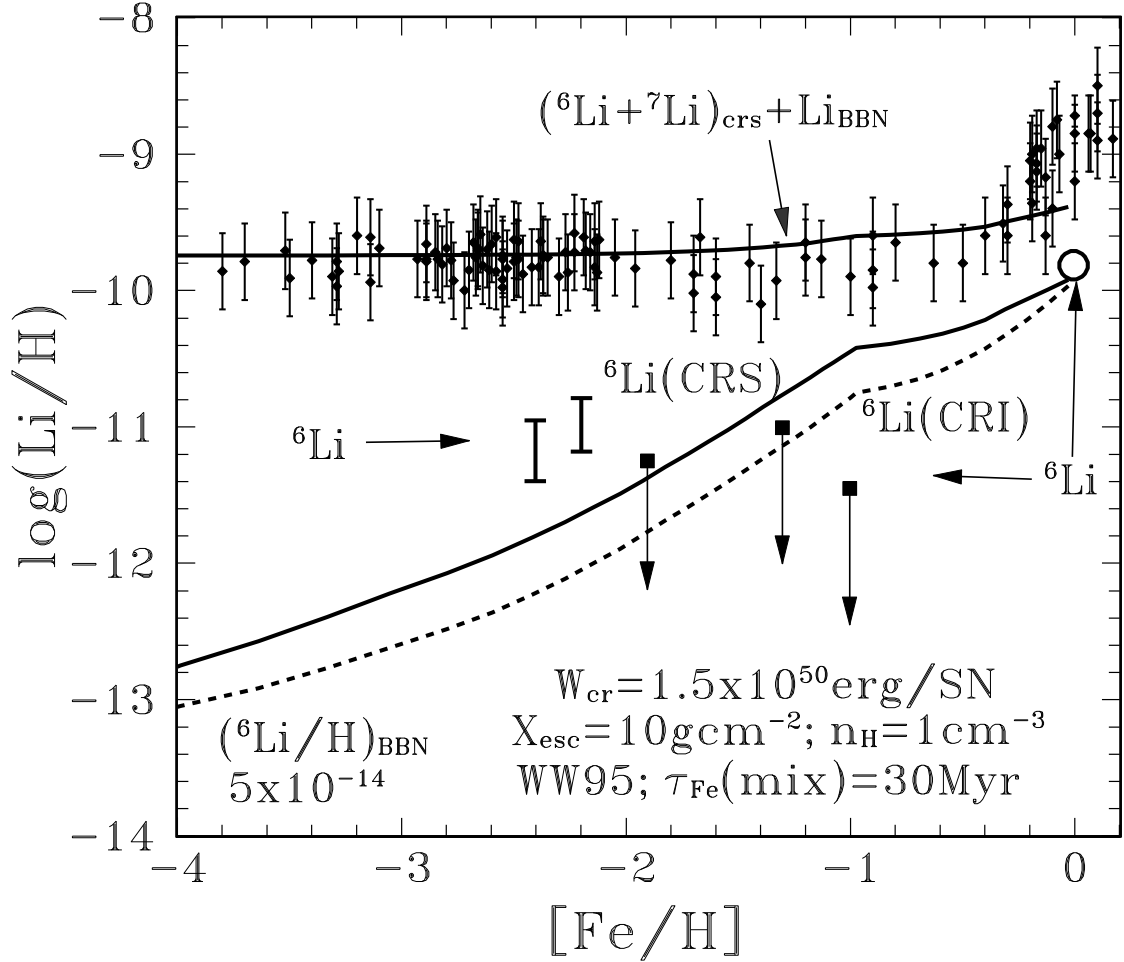}
  \end{center}
  \caption{The evolution of the total Li and $^6$Li. The $^6$Li data
  at [Fe/H] of $-$2.4, $-$2.2, $-$1.9, $-$1.3 and
$-$1 are from the summary of Hobbs (1999), and the meteoritic
value at [Fe/H]=0 is from Grevesse et al. (1996). The total Li
data is from a compilation by M. Lemoine (private communication
1997). The calculations are for the CRS and CRI models.}
\end{figure}

\begin{figure}[t]
  \begin{center}
    \leavevmode
\epsfxsize=15.cm \epsfbox{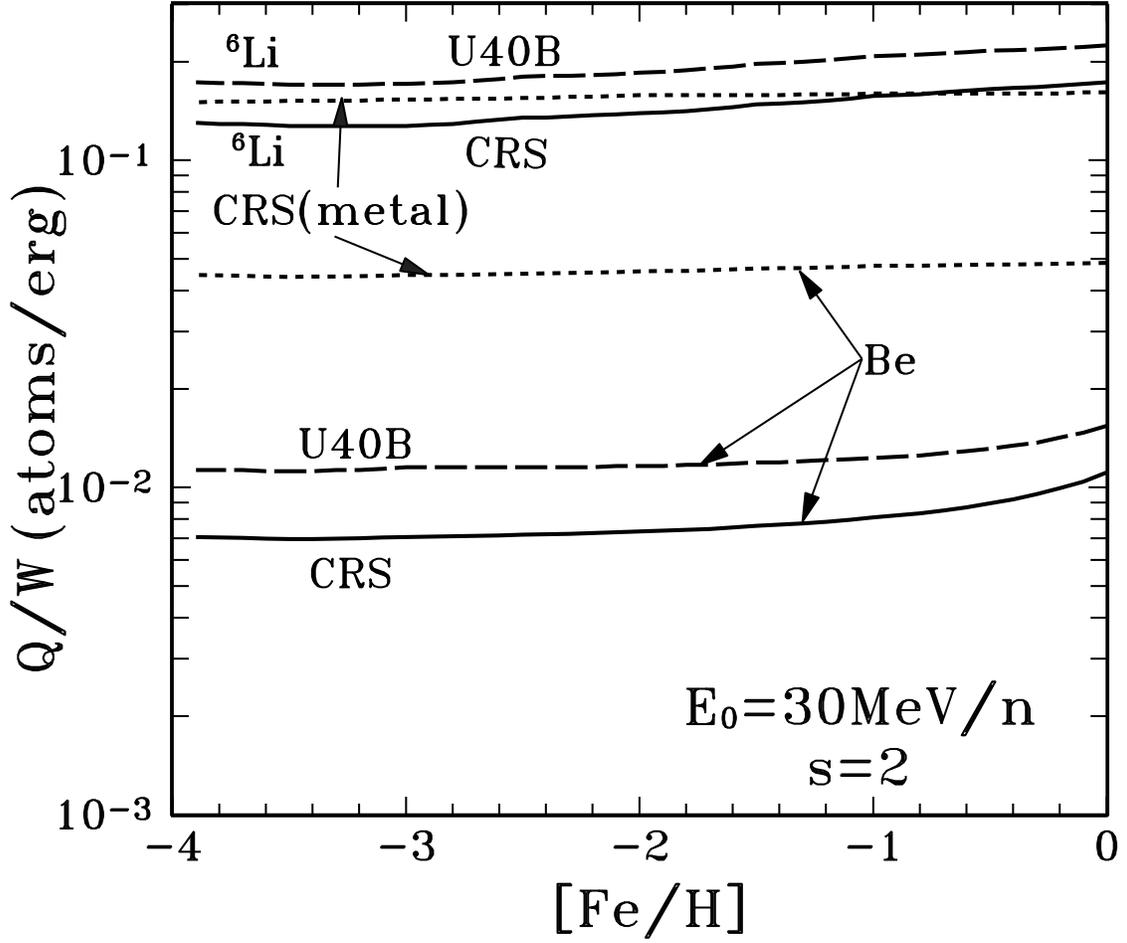}
  \end{center}
  \caption{Number of $^6$Li and Be atoms produced per unit
accelerated particle integral energy for the LECR model. The
various curves are for different accelerated particle
compositions: U40B -- the ejecta of a supernova from 40 M$_\odot$
progenitor at $10^{-4}Z_\odot$ (Woosley \& Weaver 1995); CRS --
Table~2; CRS(metal) -- identical to CRS with vanishing H and He.
Note that $Q/W$ for $^6$Li is practically independent of the
accelerated particle composition.}
\end{figure}

\begin{figure}[t]
  \begin{center}
    \leavevmode
\epsfxsize=15.cm \epsfbox{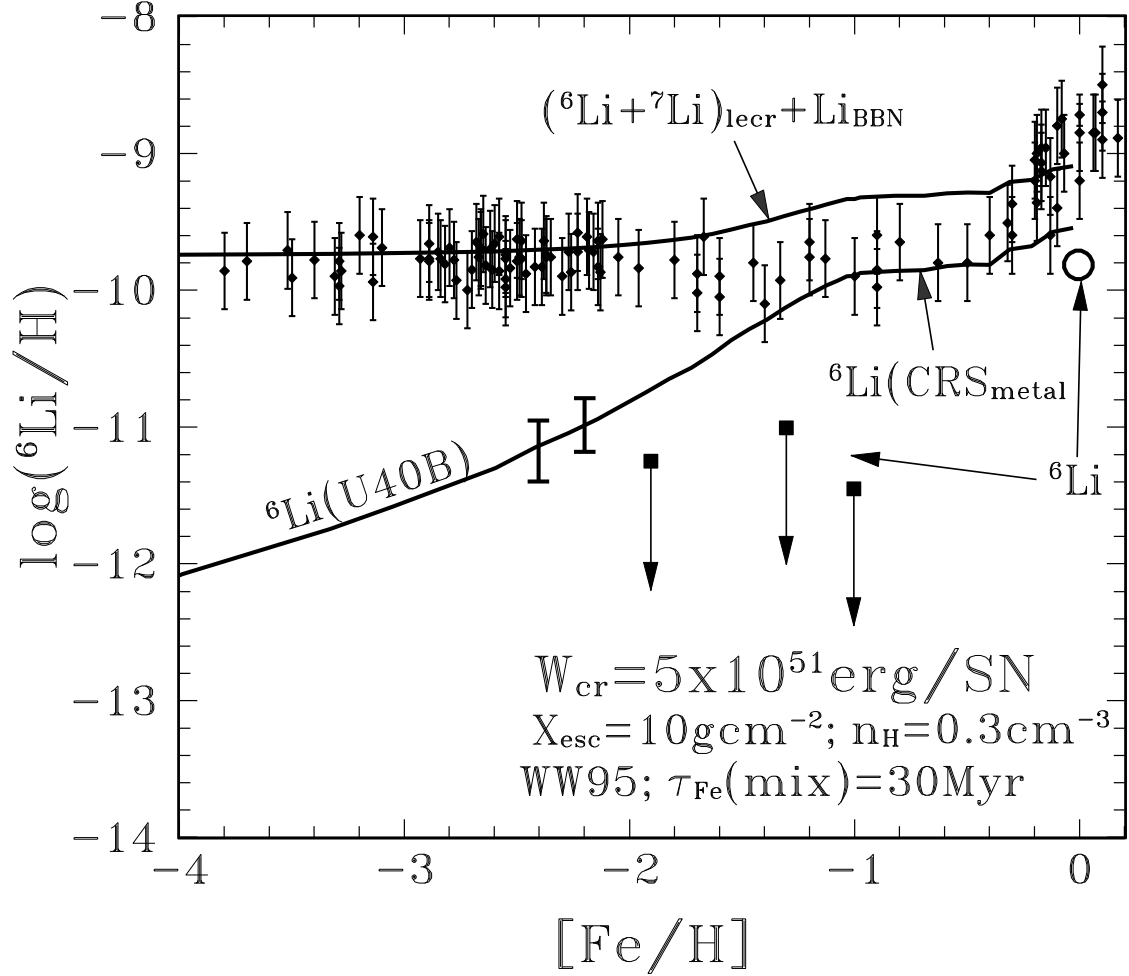}
  \end{center}
  \caption{Same as Figure~7, except that the calculations are for a
  low energy cosmic ray component originating from supernovae with
  very massive ($>$50 M$_\odot$) progenitors (Vangioni-Flam et al. 
1999).}
\end{figure}

\begin{figure}[t]
  \begin{center}
    \leavevmode
\epsfxsize=15.cm \epsfbox{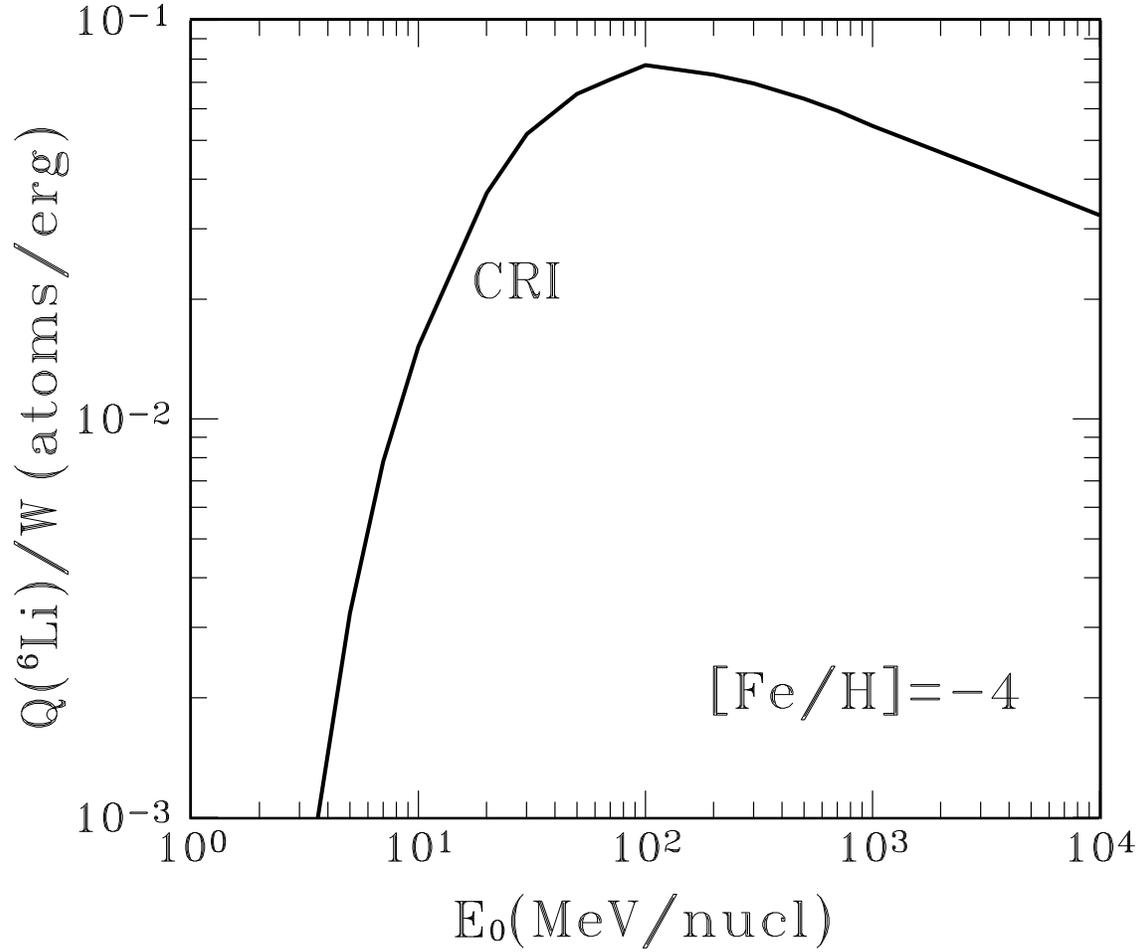}
  \end{center}
  \caption{Number of $^6$Li atoms produced per unit
integral cosmic-ray energy for the CRI composition as a function
of $E_0$ (eq.[2] with $s=2.5$) in an ambient medium of very low metallicity.}
\end{figure}

\eject

\begin{deluxetable}{llllllllll}
%\footnotesize
\tablecaption{Ejected Oxygen and Iron ($^{56}$Ni) Masses
in M$_\odot$}
%\hskip -4.0 truecm
\tablewidth{29pc}
\tablehead{ \colhead{M(progenitor)} & \colhead{ M$_{\rm
O}$(WW95)\tablenotemark{a}} & \colhead{M$_{\rm
Fe}$(WW95)\tablenotemark{a}} & \colhead{M$_{\rm
O}$(TS)\tablenotemark{b}} & \colhead{M$_{\rm
Fe}$(TS)\tablenotemark{b}} }
\startdata
12    & 0.15 & 0.054                    & 0.065 & 0.01  \\
13    & 0.18 & 0.089                    & 0.12  & 0.012 \\
15    & 0.38 & 0.064                    & 0.4   & 0.03  \\
18    & 0.77 & 0.16                     & 0.7   & 0.055 \\
20    & 1.51 & 0.09                     & 1.3   & 0.1   \\
22    & 1.73 & 0.12                     & 1.8   & 0.12  \\
25    & 2.78 & 0.20                     & 2.9   & 0.15  \\
30    & 4.07 & 5.5$\times$10$^{-7}$     & 4.5   & 0.25  \\
35    & 5.55 & 6.1$\times$10$^{-7}$     & 6.0   & 0.3   \\
40    & 6.2  & 7.3$\times$10$^{-7}$     & 8.0   & 0.35  \\
$>$40 & 6.2  & 5.5$\times$10$^{-7}$     & 12    & 0.5   \\
\enddata
\tablenotetext{a} {Woosley \& Weaver (1995)} \tablenotetext{b}
{Shigeyama \& Tsujimoto (1998); Tsujimoto \& Shigeyama (1998)}
\end{deluxetable}

\begin{deluxetable}{llllllllll}
%\footnotesize
\tablecaption{Abundances}
%\hskip -4.0 truecm
\tablewidth{15pc} \tablehead{ & \colhead{ISM} & \colhead{
CRS} & \colhead{CRI\tablenotemark{e}} }
\startdata
H        & 1                          &  1       & 1                            \\
He       & 0.084\tablenotemark{a}     &  0.135   & 0.084f(He)\tablenotemark{a}  \\
$^{12}$C & 0.00035\tablenotemark{b}   &  0.0046  & 0.00053\tablenotemark{b}     \\
$^{13}$C & 0.000004\tablenotemark{b}  &  0.00005 & 0.000006\tablenotemark{b}    \\
N        & 0.000093\tablenotemark{b}  &  0.00028 & 0.00017\tablenotemark{b}     \\
O        & 0.00074\tablenotemark{c}   &  0.0057  & 0.0015\tablenotemark{c}      \\
Ne       & 0.00012\tablenotemark{d}   &  0.00063 & 0.0003                       \\
Mg       & 0.000038\tablenotemark{d}  &  0.0011  & 0.00076\tablenotemark{d}     \\
Si       & 0.000036\tablenotemark{d}  &  0.0011  & 0.00072\tablenotemark{d}     \\
S        & 0.000016\tablenotemark{d}  &  0.00014 & 0.00008\tablenotemark{d}     \\
Fe       & 0.000032\tablenotemark{b}  &  0.0011  & 0.00064\tablenotemark{b}     \\
\enddata
\tablenotetext{a} {scaled with a factor varying from 1 to 0.75
for 0$>$[Fe/H]$>$$-$4}
\tablenotetext{b} {scaled with 10$^{\rm [Fe/H]}$}
\tablenotetext{c} {scaled with 10$^{\rm 0.63[Fe/H]}$}
\tablenotetext{d} {scaled with 10$^{\rm [Fe/H]}$ for [Fe/H]$>-1$;
    3$\times$10$^{\rm [Fe/H]}$ for [Fe/H]$<-1$}
\tablenotetext{e} {at all [Fe/H] except [Fe/H]=0 where the CRI
abundances are the same as those of the CRS model}
\end{deluxetable}

\end{document}